\documentclass[superscriptaddress,showpacs,aps,twocolumn]{revtex4}
\usepackage{amssymb}
\usepackage{amsmath}
\usepackage{graphicx}
\usepackage{color}
\usepackage{amsfonts}

\definecolor{red}{rgb}{1,0,0}
\begin{document}

\title{Trajectory-dependent energy loss for swift He atoms axially scattered
off a silver surface}
\author{C. A. R\'{\i}os Rubiano}
\affiliation{Instituto de Astronom\'{\i}a y F\'{\i}sica del Espacio (CONICET-UBA),
Casilla de correo 67, sucursal 28, 1428 Buenos Aires, Argentina.}
\author{G.A. Bocan}
\affiliation{Centro At\'{o}mico Bariloche, Comisi\'{o}n Nacional de Energ\'{\i}a At\'{o}%
mica, and Consejo Nacional de Investigaciones Cient\'{\i}ficas y T\'{e}%
cnicas, S.C. de Bariloche, R\'{\i}o Negro, Argentina.}
\author{J.I. Juaristi}
\affiliation{Departamento de F\'{\i}sica de Materiales, Facultad de Qu\'{\i}micas,
UPV/EHU, 20018 San Sebasti\'{a}n, Spain.}
\affiliation{Donostia International Physics Center (DIPC) and Centro de F\'{\i}sica de
Materiales CFM/MPC (CSIC-UPV/EHU), 20018 San Sebasti\'{a}n, Spain.}
\author{M.S. Gravielle\thanks{%
Author to whom correspondence should be addressed.\newline
Electronic address: msilvia@iafe.uba.ar}}
\affiliation{Instituto de Astronom\'{\i}a y F\'{\i}sica del Espacio (CONICET-UBA),
Casilla de correo 67, sucursal 28, 1428 Buenos Aires, Argentina.}

\begin{abstract}
Angle- and energy-loss- resolved distributions of helium atoms grazingly
scattered from a Ag(110) surface along low indexed crystallographic
directions are investigated considering impact energies in the few keV
range. Final projectile distributions are evaluated within a semi-classical
formalism that includes dissipative effects due to electron-hole excitations
through a friction force. For mono-energetic beams impinging along the $[1%
\bar{1}0]$, $[1\bar{1}2]$ and $[1\bar{1}0]$ directions, the model predicts
the presence of multiple peak structures in energy-loss spectra. Such
structures provide detailed information about the trajectory-dependent
energy loss. However, when the experimental dispersion of the incident beam
is taken into account, these energy-loss peaks are completely washed out,
giving rise to a smooth energy-loss distribution, in fairly good agreement
with available experimental data.
\end{abstract}

\maketitle
\date{\today}

\section{INTRODUCTION}

During the last decade many experimental \ and theoretical works were
devoted to study the energy lost by atomic projectiles after grazingly
colliding \ with crystal surfaces under axial surface channeling conditions,
i.e. impinging along low indexed crystallographic directions \cite%
{Robin0305,Garcia06,Valdes08,Chen11}. Recently the subject has attracted
renewed attention as a consequence of the experimental observation of fast
atom diffraction (FAD) from crystal surfaces \cite{Schuller07,Rousseau07},
where energy loss processes are considered to play an important role against
quantum coherence \cite{Winter11}. At present FAD phenomena have been
observed for a wide variety of materials, including insulators \cite%
{Schuller07,Rousseau07}, metals \cite{Bundaleski08,Busch09} and
semi-conductors \cite{Khemliche09}, as well as adsorbate covered metal
surfaces \cite{Schuller09} and ultrathin films \cite{Seifert10} and organic
molecules \cite{SeifertPRL13} on metal substrates. In all these cases the
preservation of the quantum coherence becomes a crucial factor for the
observation of the diffraction patterns. Since the energy transfer from the
projectile to the surface is considered a relevant source of decoherence in
FAD, in recent articles \cite{Lienemann11,Busch12} the energy lost by fast H
and He atoms scattered from a LiF($001$) surface under axial surface
channeling conditions was experimentally investigated by recording the
angular distribution of scattered projectiles in coincidence with their
energy loss. For H projectiles it was found that inelastic electronic
processes are responsible for the diffuse background present in experimental
FAD spectra, while for He impact contributions of surface electronic
excitations were found to be significantly smaller, strongly reducing the
presence of the inelastic background in the corresponding angular
distributions. On the other hand, in Ref. \cite{Lienemann13}
trajectory-dependent energy loss for He atoms grazingly colliding with a LiF(%
$001$) surface along low indexed directions was experimentally and
theoretically studied \ considering a perpendicular energy, associated with
the motion normal to the surface plane, higher than the perpendicular energy
range for FAD.

Since for metals\ the absence of an energy threshold for electronic
excitations favors the projectile energy dissipation, in this article we
study the energy loss distribution for swift He atoms impinging on a metal
surface - Ag($110$) \ - under the same conditions for which FAD patterns
have been reported \cite{Bundaleski08,Rubiano13}. Precisely, this collision
system corresponds to the first and simplest metallic case for which FAD
effects were experimentally observed, in conjunction with measurable energy
losses \cite{Bundaleski08,KhemlicheNIMB09,Bundaleski11}. The projectile
energy loss due to electronic transitions from the metal\ surface is here
described by means of a semi-classical formalism that takes into account the
energy dissipation along different classical paths without including effects
of quantum coherence. The influence of quantum interferences in the
projectile energy-loss\ spectrum is expected to be minor because even for
insulator surfaces, where valence electrons are tighter than in metals,
coherence quantum effects are completely washed out when partial
contributions coming from different initial crystal states are added to
obtain the transition probability to a given final state \cite{Gravielle07}.

To evaluate the energy lost by axially channeled He atoms we introduce a
friction force in \ Newton%
\'{}%
s equations for the projectile trajectory. The friction force is expressed
in terms of the transport cross section at the Fermi level, corresponding to
the screened potential of the atom embedded in an electron gas \cite%
{Juaristi08}. At each point of the trajectory we use a local electronic
density that is evaluated from an accurate density-functional theory (DFT)
calculation. Both the potential \ for He-Ag($110$) and the surface
electronic density are evaluated on equal footing, i.e. from DFT
calculations within the same conditions. The projectile-surface interaction
is represented by a potential energy surface (PES) that was built from a
large set \ of \textit{ab initio} data obtained with the DFT-based
\textquotedblleft \textsc{Quantum Espresso}\textquotedblright\ code \cite%
{Giannozzi09}, combined with a sophisticated interpolation technique \cite%
{CRP}. From such \textit{ab initio} values we derived a three-dimensional
(3D) PES, taking into account the projectile's three degrees of freedom. No
average of the surface potential nor the electronic density along the
incidence direction was considered in the calculation. In Ref. \cite%
{Rubiano13} the quality of such a PES \ was tested by means of FAD patterns
for perpendicular energies in the range $0.1$ eV- $0.5$ eV.

Double differential - angle and energy-loss resolved - probabilities for He
atoms scattered along three different crystallographic directions - $[1\bar{1%
}0]$, $[1\bar{1}2]$, and $[001]$ - are analyzed, considering different
incidence energies and angles. The work is organized as follows. The
theoretical method is summarized in Sec. II, results are presented and
discussed in Sec. III, and in Sec. IV we outline our conclusions. Atomic
units (a.u.) are used unless otherwise stated.\medskip

\section{THEORETICAL MODEL}

The final projectile distribution originated by inelastic collisions with
the surface is derived from classical trajectory calculations by including
the energy lost by the atom along the classical path. For grazing incidence
of He atoms with energies in the few keV range, electron-hole pair
excitations represent the main mechanism of projectile energy loss, while
contributions of nuclear scattering are expected to be negligible \cite%
{Robin01,Lederer07}. Due to the fact that in a metal there is no minimum
energy required to excite electron--hole pairs, for atoms moving with
velocities lower than the Fermi velocity of the metal, the electronic
stopping power, i.e. the energy lost per unit path length, has a linear
dependence on velocity \cite{Ritchie59}. Then, the dissipative force
experienced by the moving atom can be expressed as $\mathbf{F}_{\mathrm{diss}%
}=-\mu \mathbf{v}$ , where $\mu $ is a friction coefficient and $\mathbf{v}$
is\textbf{\ }the velocity of the atom. Here the coefficient $\mu $ is
calculated within the Local Density Friction Approximation \ (LDFA) \cite%
{Juaristi08}. This model has been successfully applied to study dissipative
effects of atoms and molecules interacting with different metal surfaces 
\cite{Goikoetxea09,Martin12,Alducin13}.

Within the LDFA the modulus of the dissipative force acting on the
projectile is calculated in terms of the transport cross section at the
Fermi level $\sigma _{\mathrm{tr}}(k_{F})$ as \cite{Echenique81}: 
\begin{equation}
F_{\mathrm{diss}}=n_{0}v\ k_{F}\ \sigma _{\mathrm{tr}}(k_{F}),
\label{fdis-tot}
\end{equation}%
where $n_{0}$ is the electron gas density and $k_{F}$ is the corresponding
Fermi momentum. At each position $\mathbf{R}$ along the classical trajectory
the electron gas density $n_{0}$ is approximated by the local electronic
density $n(\mathbf{R})$, which is evaluated from \textit{ab} \textit{initio}
calculations and within the same conditions as the PES. Then, the friction
coefficient $\mu (\mathbf{R})$ is expressed in terms of the transport cross
section at the Fermi level associated with the electron scattering at the
potential induced by the He projectile in the electron gas \cite{Echenique81}%
. Such a potential is evaluated using DFT \cite{Zaremba77}. In this way, the
model includes nonlinear effects both in the medium response to the atomic
potential (nonlinear screening) and in the calculation of the relevant
cross-sections for the energy loss process.\ 

The classical trajectory of the projectile is obtained by solving Newton%
\'{}%
s equations \cite{Juaristi08,Goikoetxea09}:

\begin{equation}
m_{P}\frac{d^{2}\mathbf{R}}{dt^{2}}=-\mathbf{\nabla }V_{SP}(\mathbf{R})-\mu (%
\mathbf{R})\frac{d\mathbf{R}}{dt},  \label{newton}
\end{equation}%
where $m_{P}$ is the projectile mass and $V_{SP}(\mathbf{R})$ is the
projectile-surface potential. The first term on the right side of Eq. (\ref%
{newton}) is the adiabatic force obtained from the 3D PES, while the second
term is the dissipative force experienced by the atom. From the solutions of
Eq. (\ref{newton}) for different initial positions on the surface plane, the
final projectile distribution $dP/dE_{f}d\Omega _{f}$ is obtained by
counting the number of classical paths ending with final momentum $\mathbf{K}%
_{f}$ \ in the direction of the solid angle $\Omega _{f}\equiv (\theta
_{f},\varphi _{f})$ and final energy $E_{f}=K_{f}^{2}/(2m_{p})$, \ where $%
\theta _{f}$ and $\varphi _{f}$ \ are the final polar and azimuthal angles,
respectively, with $\varphi _{f}$ measured with respect to the incidence
direction in the surface plane.

The interaction energy of the He atom with the Ag($110$) surface is
described with a full adiabatic 3D PES that depends on the atomic position $%
\mathbf{R}=(X,Y,Z)$. The PES is constructed by interpolating \cite{CRP} over
a grid of \textit{ab initio} energies for 42 equidistant points $Z$, ranging
from the asymptotic region to 2 a.u. below the topmost atomic layer, and 6 $%
(X,Y)$ sites uniformly spread on the surface unit cell. All \textit{ab initio%
} data are obtained from the DFT-based \textquotedblleft \textsc{Quantum
Espresso}\textquotedblright\ code ~\cite{Giannozzi09}. Details regarding the
PES calculation can be found in Ref. \cite{Rubiano13}.

\section{RESULTS}

In this work we extend a previous study \cite{Rubiano14} to investigate the
energy lost by $^{3}$He atoms grazingly colliding with Ag($110$) \ along the 
$[1\bar{1}0]$, $[1\bar{1}2]$ and $[001]$ channels. Impact energies $E_{i}$
ranging from $0.5$ keV to $2$ keV and perpendicular energies $E_{i\bot
}=E_{i}$ $\sin ^{2}\theta _{i}$ varying from $0.15$ eV to $0.87$ eV are
considered, $\theta _{i}$ being the incidence angle measured with respect to
the surface plane. Energy- and angle- resolved distributions of
inelastically scattered He atoms are classically derived by solving Eq. (\ref%
{newton}) for $4\times 10^{5}$ random initial positions that vary within a
surface area equal to $4\times 4$ unit cells. For all the trajectories the
initial atom-surface distance is chosen equal to the lattice constant,
corresponding to a region where the surface interaction is completely
negligible \cite{Rubiano13}. The differential probability $dP/dE_{f}d\Omega
_{f}$ is calculated by considering a dense grid of $E_{f}$, $\theta _{f}$,
and $\varphi _{f}$ values ($100\times 100\times 100$), which is used to
build the cells where final momenta $\mathbf{K}_{f}$ are assigned. The
energy-loss distribution $dP/d\omega $, as a function of the lost energy $%
\omega =$ $E_{i}-$ $E_{f}$, is straightforwardly derived from $%
dP/dE_{f}d\Omega _{f}$ \ by integrating on the solid angle $\Omega _{f}$.
Fig. 1\ (a) shows the friction coefficient $\mu $ for the He atom moving in
an electron gas of local density $n_{0}$, as a function of the mean electron
radius $r_{s}$, defined as $r_{s}=[3/(4\pi n_{0})]^{1/3}$. \ Electron
density contours for two different $Z$- distances to the surface are plotted
in Fig. 1 (b).

In Fig. 2 we show $dP/d\omega $ for $E_{i}=1$ keV and $\theta
_{i}=1.0^{\circ }$ (that is, $E_{i\bot }=0.30$ eV) considering the incidence
directions $[1\bar{1}0]$, $[1\bar{1}2]$ and $[001]$. These energy-loss
distributions, obtained without including the experimental spread of the
incident beam in the calculations, present well defined structures with
multiple peaks, \ resembling the energy-loss spectra reported in Ref. \cite%
{Moix10}. In order to identify the origin of these peaks, labeled with
letters A, B, and C in Figs. 2 a), b) and c) respectively, representative
trajectories contributing to them are plotted in Fig. 3. For the three
incidence directions we find that every energy-loss peak is related to a
defined set of projectile trajectories. For incidence along the channel $[1%
\bar{1}0]$ (Fig. 2 (a)) the peak A$_{1}$, which presents the highest
intensity, is associated with trajectories that suffer strong azimuthal
deviations with respect to the incidence direction, corresponding to
classical rainbow scattering. Such paths probe regions with low electron
density, producing the lowest energy loss of the spectrum. Instead, the
trajectories that contribute to the peak A$_{3}$ correspond to He atoms that
move over the rows of topmost Ag atoms that form the channel, suffering
almost no deviation. Even though these projectiles are the ones that least
approach the surface as they do not enter the channel, they probe the region
with the highest electron density, thus suffering the highest energy loss.
Lastly, two different kind of trajectories contribute to the peak A$_{2}$:
one of them running parallel to the channel in the middle position between
rows and the other running far away from the surface plane and suffering an
azimuthal deflection.

A similar structure, with three peaks, is observed for incidence along the $%
[001]$ channel (Fig. 2 (c)), but in this case, the peak C$_{2}$ is the one
associated with rainbow scattering, while the peaks C$_{1}$ and C$_{3}$ \
are related to He atoms that move in the middle or on top of the channel
rows, respectively. On the other hand, the energy-loss distribution for
incidence along the $[1\bar{1}2]$ channel (Fig. 2 (b)) displays a completely
different structure, with only two peaks - B$_{1}$ and B$_{2}$\ - at the
extremes of the distribution. These peaks are associated with projectiles
running in the middle of the channel or over the rows of topmost surface
atoms, respectively. We have thus found that it is the dependence of energy
loss on the kind of He trajectory that determines the multiple-peak
structure in energy-loss distributions for mono-energetic incident beams
(named here primary distributions). This result could be used to study the
effective corrugation of the surface electronic density. However, the
experimental determination of these primary energy-loss spectra would
require an energy resolution better than one eV, which is not yet reachable
with present experimental capabilities.

With the aim of comparing the energy-loss spectra with available
experimental data \cite{Bundaleski08,Bundaleski11,note-dir0}, in Fig. 4 we
plot \ the energy-loss distribution obtained by including the energy profile
of the experimental incident beam through convolution \cite{note-conv}. Two
different initial conditions are considered: (a) $E_{i}=0.5$ keV, $\theta
_{i}=1.5^{\circ }$ (i.e. $E_{i\bot }=0.34$ eV) and (b) $E_{i}=1$ keV, $%
\theta _{i}=1.0^{\circ }$ (i.e. $E_{i\bot }=0.30$ eV). In both cases, when
the experimental energy dispersion is introduced in the calculation we
obtain a smooth energy-loss curve, without any signature of the
primary-energy loss structure. For $E_{i}=0.5$ keV the experimental curve is
fairly well reproduced by the simulation, but the agreement deteriorates
when the energy increases and for $E_{i}=1$ keV \ the maximum of the energy
loss distribution overestimates the experimental value. Primary
distributions of Fig. 2 present completely different shapes but, for
incidence along the $[1\bar{1}2]$ and $[001]$ channels they produce similar
convoluted energy-loss spectra, with a mean energy loss slightly higher than
the one obtained for the $[1\bar{1}0]$ direction, as observed in Fig. 4.
This last result is unexpected at first glance because the average
electronic density probed by He atoms that move over the atomic rows of the $%
[1\bar{1}0]$ channel is higher than the one corresponding to any of the
other channels. However, since projectiles running along the $[1\bar{1}2]$
or $[001]$ directions suffer a smaller friction in the incoming path than in
the case of the $[1\bar{1}0]$ direction, they can get closer to the surface
thus loosing more energy in the whole trajectory. This interplay between the
distance of maximum approach to the surface and the energy lost along the
path is also observed for the different perpendicular energies considered in
this work, as shown in Fig. 7.

In relation to the angular distribution of inelastically scattered He atoms,
in all the cases it shows the usual banana shape \cite{Meyer95}, with final
polar and azimuthal angles lying inside a circular annulus, characteristic
of the axial channeling conditions. Due to the fact that in our model the
friction force acts along the direction of the velocity, the energy loss
affects mainly the component of the projectile velocity along the incidence
channel. Then, initial and final energies associated with the motion normal
to the channel are very similar, almost strictly verifying $E_{f}\ (\varphi
_{f}^{2}+\theta _{f}^{2})\simeq E_{i}\ \theta _{i}^{2}$. \ Double
differential probabilities, $d^{2}P/d\omega d\varphi _{f}$ , as a function
of the lost energy $\omega $ and the azimuthal angle $\varphi _{f}$, are
displayed in Fig. 5 for the incidence conditions of Fig. 2. For the three
channels the two-dimensional (2D) angle- and energy-loss- resolved
distributions present a double peak energy-loss structure for most angles,
with sharp maxima at the outermost azimuthal angles, which are associated
with classical rainbow dispersion. The shape of these \ 2D spectra strongly
varies with the incidence direction, showing energy-loss peaks associated
with rainbow scattering only for the $[1\bar{1}0]$ and $[001]$ channels. \
Notice that while energy-loss structures give information about the
electronic density along the channel, the angular positions of rainbow peaks
depend on the corrugation of the surface potential across the incidence
direction \cite{Schuller04,Schuller07b,Tiwald10,Gravielle13}. Thus,
experiments in coincidence, measuring simultaneously angular and energy-loss
distributions with a certainly high energy resolution, might provide useful
insights on the atom-surface interaction.

Finally, to investigate the dependence of the energy-loss on $E_{i\bot }$,
in Fig. 6 we display primary energy-loss distributions for $1$ keV He \
atoms, as a function of the normalized lost energy $\omega _{\text{{\small %
norm}}}=\omega \ /\left\langle \omega \right\rangle $, where $\left\langle
\omega \right\rangle $ denotes the mean energy loss, considering different
perpendicular energies. In the figure we have labeled as \textit{top}, 
\textit{middle} and \textit{rb} the peaks associated with trajectories
running along top or middle-channel rows, or contributing to the rainbow
angle, respectively. We found that the relative positions of the energy-loss
peaks change with the perpendicular energy. Such an energy displacement is
more notorious for the directions $[1\bar{1}2]$ and $[001]$, which present a
high corrugation of the electronic density along the channel. \ While for
the $[1\bar{1}0]$ direction \ the type of trajectories contributing to each
peak do not vary with $E_{i\bot }$ in the considered range, for the other
two directions \textit{top} (\textit{middle}) trajectories contribute to the
region of low (high) or high (low) lost energies, depending on the
perpendicular energy. This fact is again related to the strong corrugation
of the electronic density along these channels. In addition, the mean energy
loss for a given incidence direction, normalized with the impact energy $%
E_{i}$, shows a linear behavior as a function of the perpendicular energy,
increasing as $E_{i\bot }$ augments, as observed in Fig. 7. In all the
cases, mean energy loss values for the channels $[1\bar{1}2]$ and $[001]$
are similar, being higher than the ones corresponding to the $[1\bar{1}0]$
channel, as discussed above.

\section{CONCLUSIONS}

We have studied the energy lost by helium atoms after grazingly colliding
with a silver surface along low indexed crystallographic directions. The
distribution of inelastically scattered atoms was obtained within a
semi-classical formalism that incorporates a friction force in the classical
dynamics equations allowing\ for the calculation of the trajectory-dependent
energy loss. For the $[1\bar{1}0]$ , $[1\bar{1}2]$ and $[001]$ channels we
found that\ energy-loss distributions corresponding to mono-energetic
incidence beams display well defined structures, with several sharp maxima
that are related to trajectories that probe regions with different densities
and potential energies. However, these energy-loss structures are completely
blurred out by the experimental spread of the incident beam, which produces
a broad energy-loss distribution with only one maximum. Even though these
last distributions are in fairly good agreement with available experimental
data \cite{Bundaleski08,Bundaleski11}, experiments in coincidence measuring
simultaneously angle- and energy-loss- resolved spectra with a high energy
resolution would be necessary to shed light on the findings of the present
work.

\bigskip

\begin{acknowledgments}
Authors are kindly grateful to Philippe Roncin for his helpful suggestions.
C.R.R and M.S.G. acknowledge financial support from CONICET, UBA, and ANPCyT
of Argentina. G.A.B. acknowledges financial support by ANPCyT and is also
thankful to Dr. H.F. Busnengo, Dr. J.D. F\"{u}hr and Dr. M.L. Martiarena
regarding the PES calculation. J. I. J. acknowledges financial support by
the Basque Departamento de Educaci\'{o}n, Universidades e Investigaci\'{o}n,
the University of the Basque Country UPV/EHU (Grant No. IT-756-13) and the
Spanish Ministerio de Ciencia e Innovaci\'{o}n (Grant No.
FIS2010-19609-C02-02)
\end{acknowledgments}

\begin{figure}[tbp]
\includegraphics[width=0.5\textwidth]{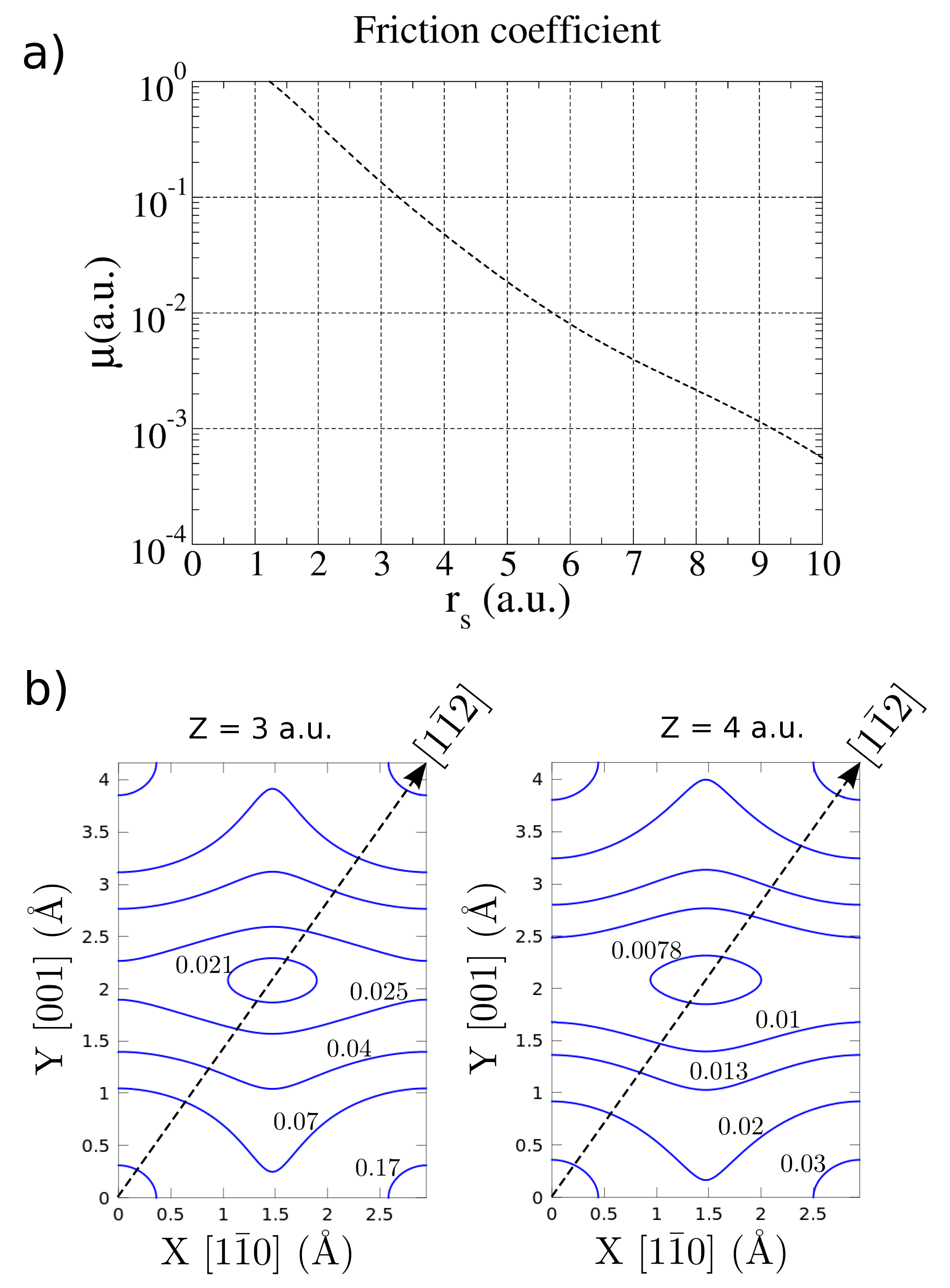}
\caption{(Color online) (a) Friction coefficient for He in an electron gas
as a function of the mean electron radius $r_{s}$. (b) Electronic density
contours for two different distances  $Z$ \ to the surface: (left) $Z=3$
a.u, (right) $Z=4$ a.u.}
\label{fig:1}
\end{figure}

\begin{figure*}[tbp]
\includegraphics[width=\textwidth]{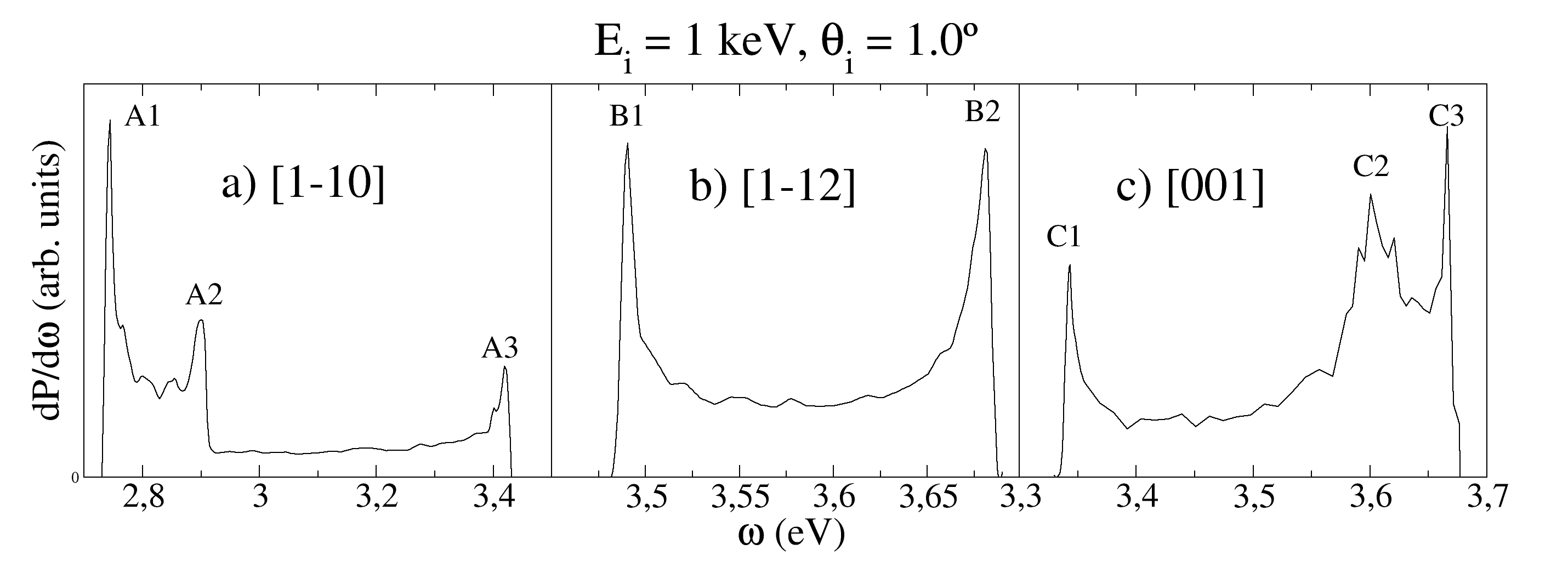}
\caption{Differential probability $dP/d\protect\omega $, as a function of
the lost energy $\protect\omega $, for He atoms impinging on a Ag(110)
surface along three different channels: (a) $[1\bar{1}0]$, \ (b) $[1\bar{1}%
2] $, and (c) $[001]$. The incidence conditions correspond to a
mono-energetic beam with $E_{i}=1$ keV and $\protect\theta _{i}=1^{{{}^{o}}}$%
. Capital letters identify the different energy-loss peaks.}
\label{fig:2}
\end{figure*}

\begin{figure}[tbp]
\includegraphics[width=0.5\textwidth]{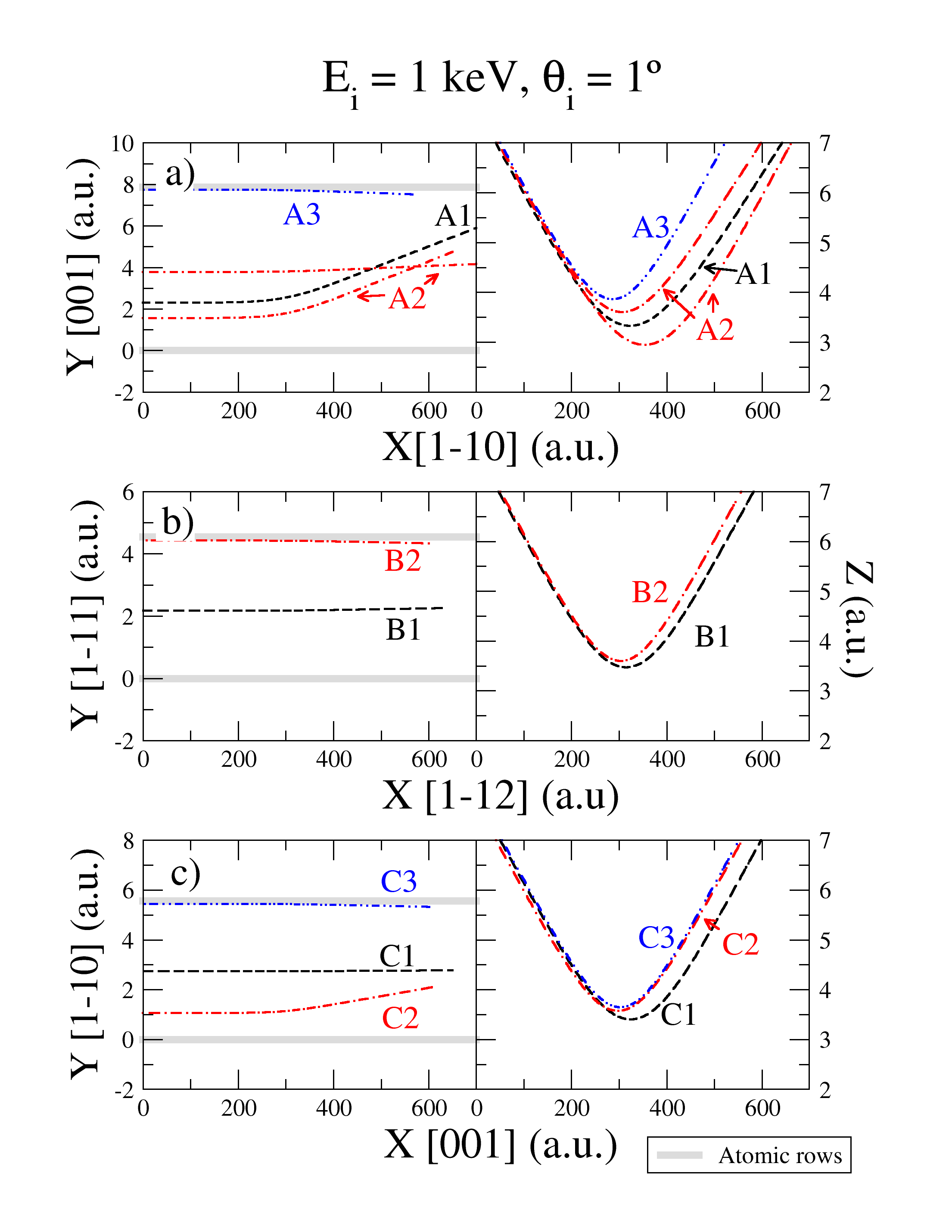}
\caption{ (Color online) For the incidence conditions of Fig. 2,
representative trajectories contributing to the different energy-loss peaks,
labeling them with the same letters as in Fig. 2. For incidence along (a) $[1%
\bar{1}0]$, (b) $[1\bar{1} 2] $, and (c) $[001]$, different trajectories are
plotted with different colors and line styles. In all the cases: left panel,
transversal position $Y$ along the path (i.e. coordinate perpendicular to
the incidence direction on the surface plane); right panel, distance $Z$ to
the topmost atomic layer along the path, both as a function of the
coordinate $X$ along the channel. Thick gray lines show positions of atomic
channel rows.}
\label{fig:3}
\end{figure}

\begin{figure}[tbp]
\includegraphics[width=0.5\textwidth]{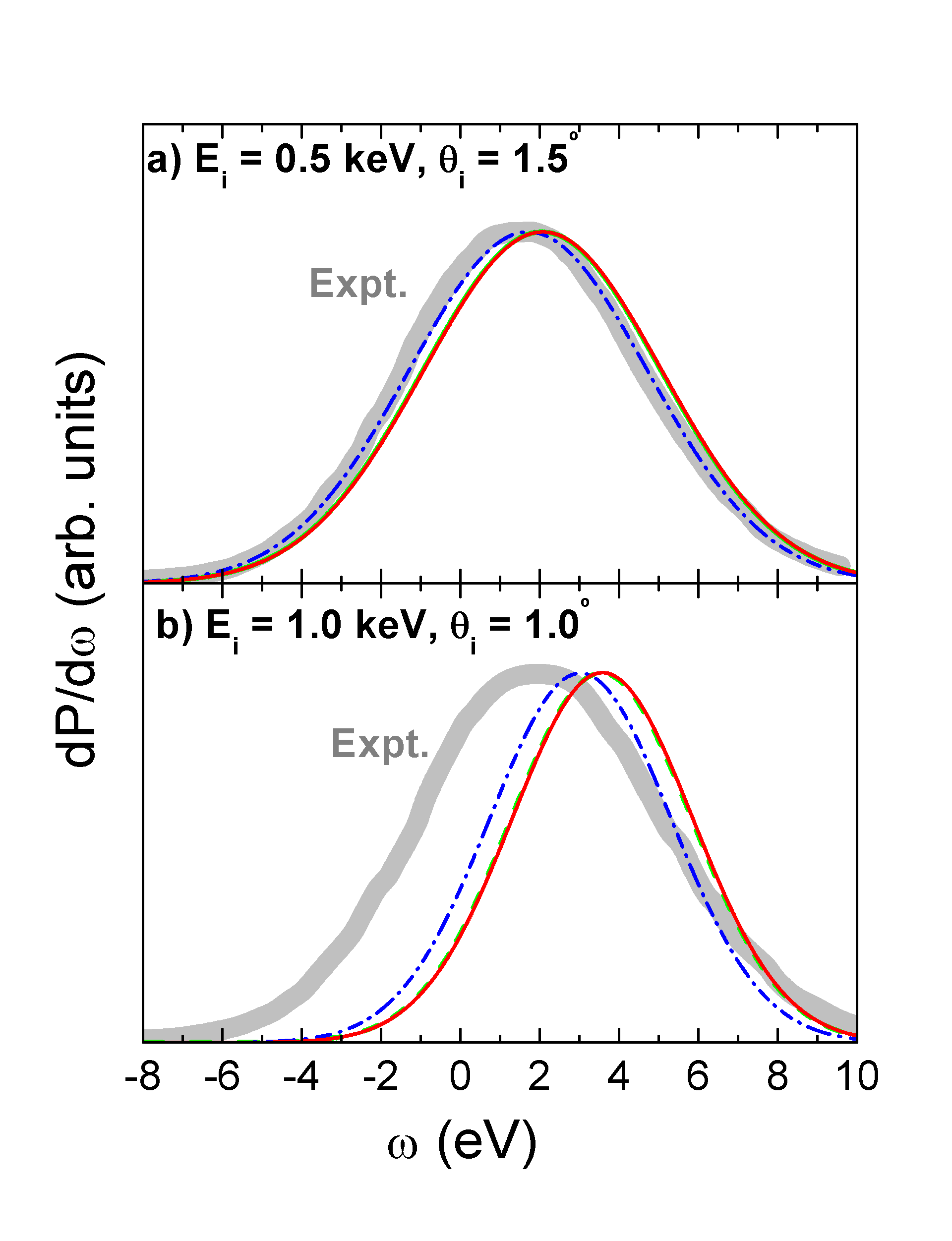}
\caption{(Color online) Energy-loss distribution, as a function of the lost
energy $\protect\omega $, for He atoms impinging on a Ag(110) surface. The
incidence conditions are (a) $E_{i}=0.5$ keV and $\protect\theta _{i}=1.5^{{%
{}^{o}}}$, and (b) $E_{i}=1$ keV and $\protect\theta_{i}=1^{{{}^{o}}}$. Red
solid, green dashed, and blue dash-dotted lines, differential probability $%
dP/d\protect\omega $, convoluted to include the experimental energy spread 
\protect\cite{note-conv}, for incidence along the $[1\bar{1}2]$, $[001]$ and 
$[1\bar{1}0]$ directions, respectively; gray solid line, experimental data
from Refs. \protect\cite{Bundaleski08,Bundaleski11,note-dir0}.}
\label{fig:4}
\end{figure}

\begin{figure}[tbp]
\includegraphics[width=0.5\textwidth]{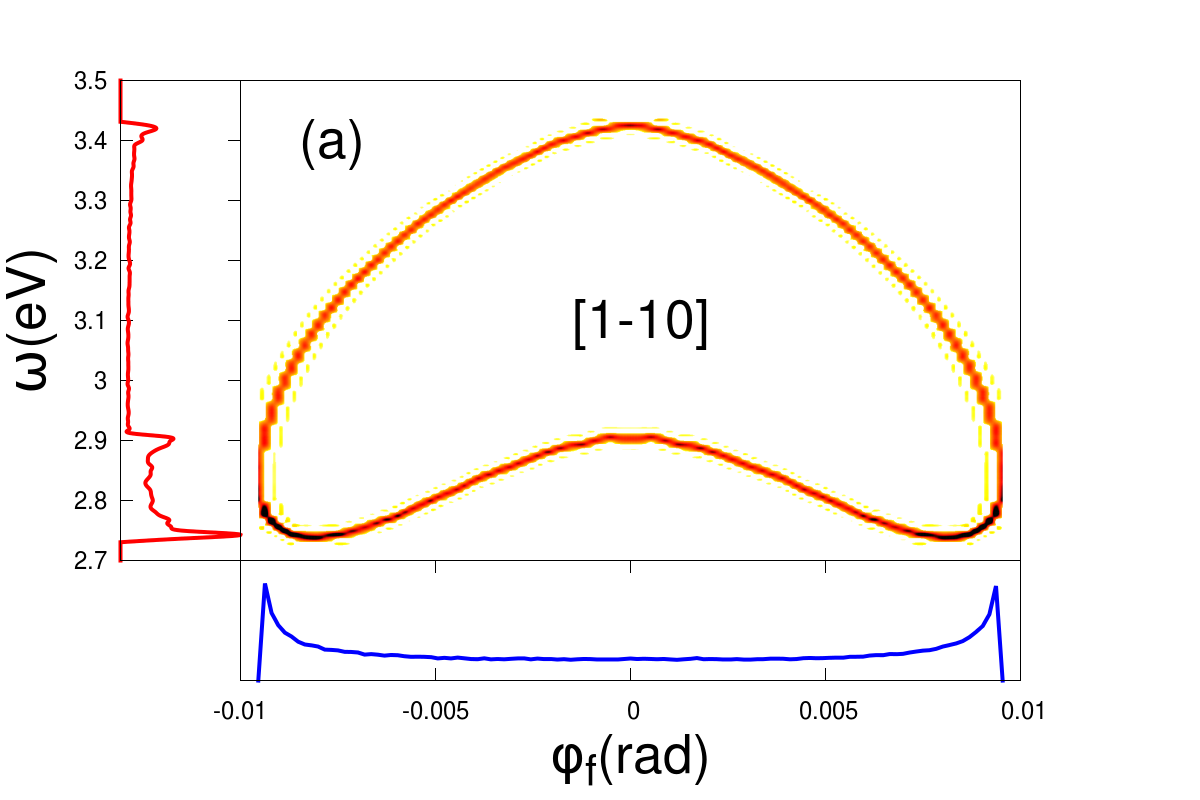} %
\includegraphics[width=0.5\textwidth]{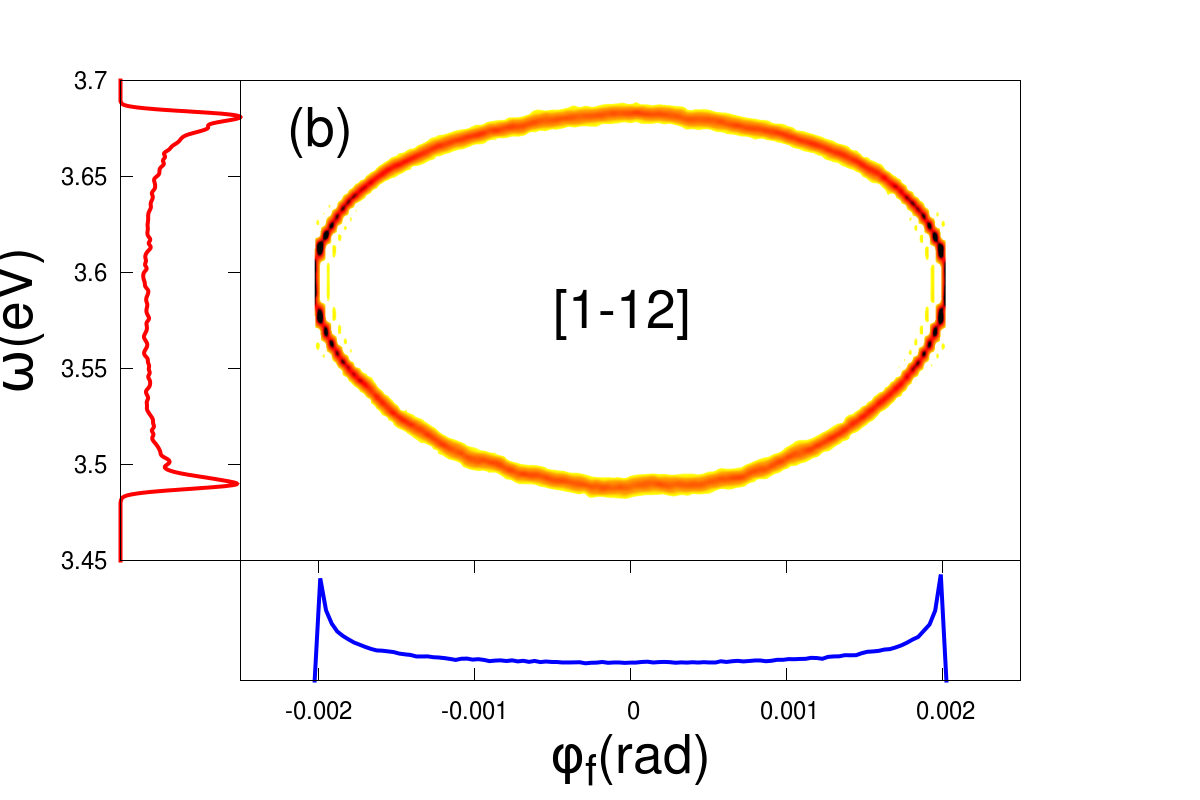} %
\includegraphics[width=0.5\textwidth]{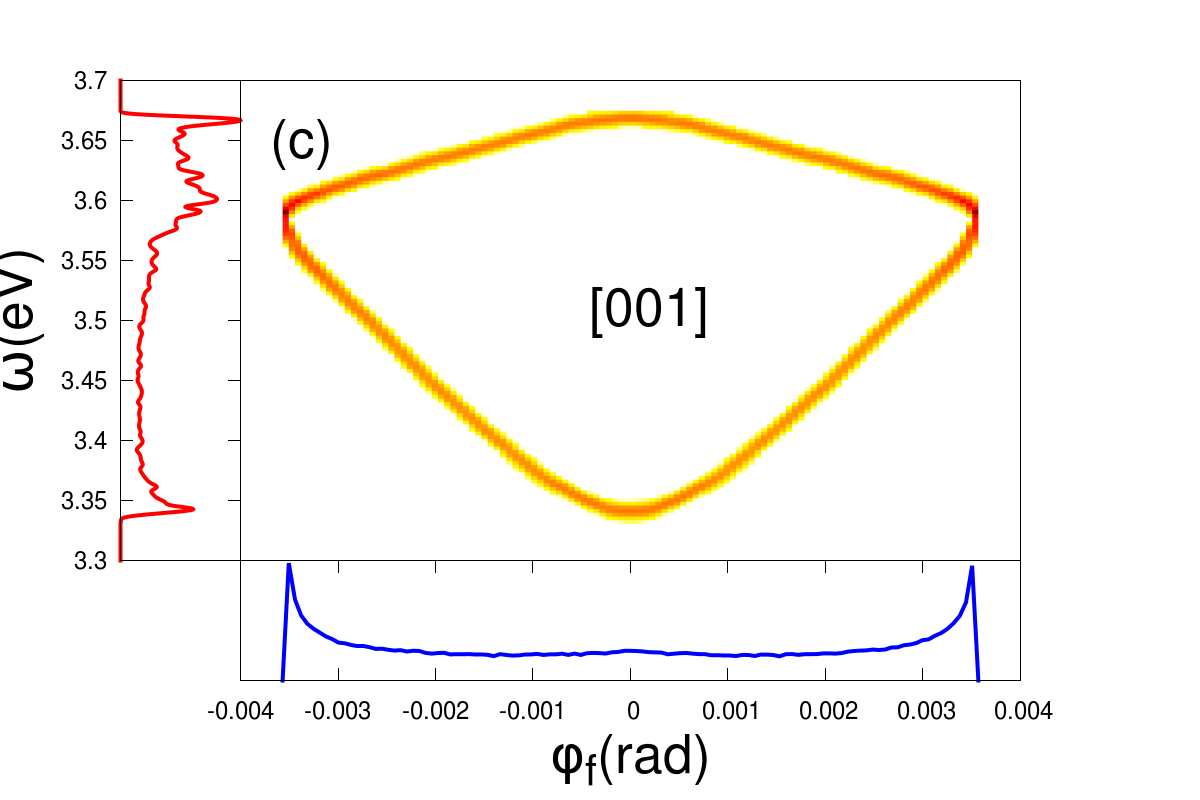}
\caption{ (Color online) 2D angle- and energy-loss distributions, as a
function of the final azimuthal angle $\protect\varphi _{f}$ and the lost
energy $\protect\omega $, for 1 keV He atoms impinging on Ag(110) with $%
\protect\theta_{i}=1^{{{}^{o}}}$. Three different incidence directions are
considered: (a) $[1\bar{1}0]$, (b) $[1\bar{1}2]$, and (c) $[001]$.
Integrated angular and energy-loss spectra are also shown in the figure.}
\label{fig:5}
\end{figure}

\begin{figure*}[tbp]
\includegraphics[width=1.0\textwidth]{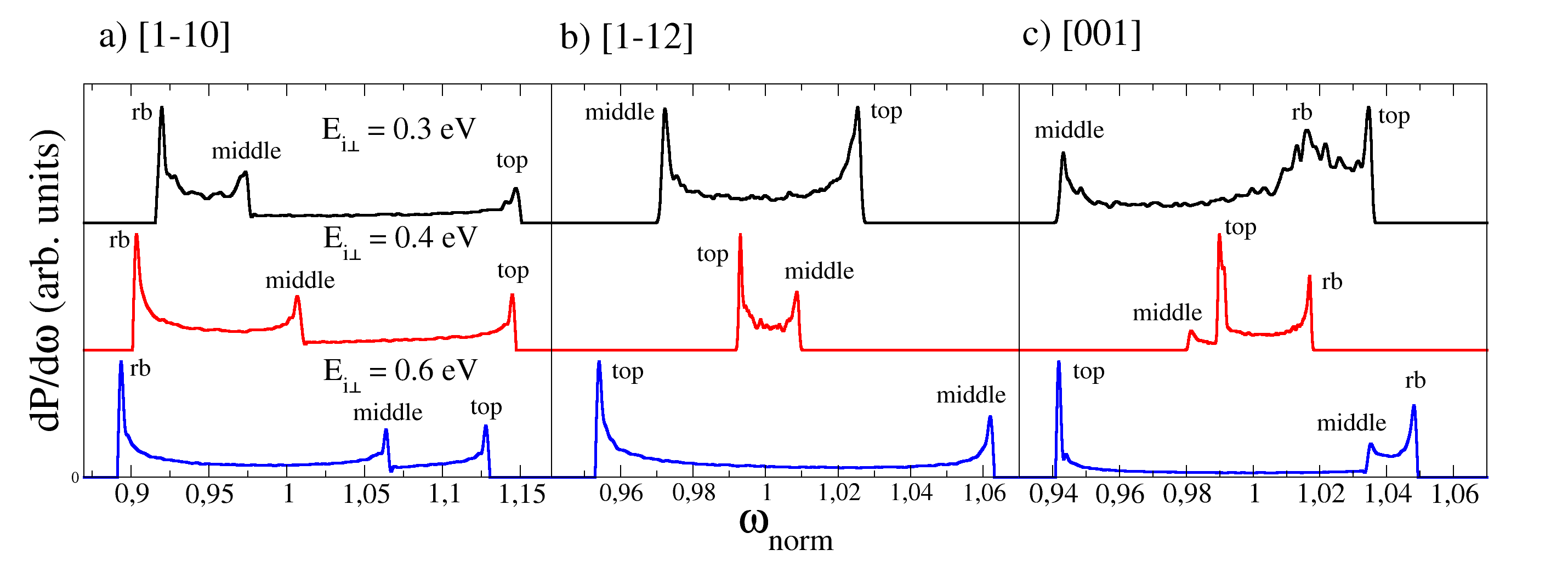}
\caption{(Color online) Energy-loss spectra, as a function of the normalized
lost energy, $\protect\omega _{norm}=\protect\omega /\langle \protect\omega %
\rangle $, for 1 keV He atoms impinging on a Ag(110) surface along three
different channels: (a) $[1\bar{1}0]$, (b) $[1\bar{1}2]$, and (c) $[001]$.
Upper black curves, energy-loss distributions for $E_{i\bot }=0.3$ eV (i.e. $%
\protect\theta _{i}=1.0^{{{}^{o}}}$); middle red curves, for $E_{i\bot }=0.4$
eV (i.e. $\protect\theta _{i}=1.2^{{{}^{o}}}$); and lower blue curves, for $%
E_{i\bot }=0.6$ eV (i.e. $\protect\theta _{i}=1.4^{{{}^{o}}}$). Labels "%
\emph{rb}", "\emph{top}", and "\emph{middle}" identify peaks associated with
paths contributing to the rainbow angle or running on top or in the middle
of the topmost atomic rows, respectively.}
\label{fig:6}
\end{figure*}

\begin{figure}[tbp]
\includegraphics[width=0.5\textwidth]{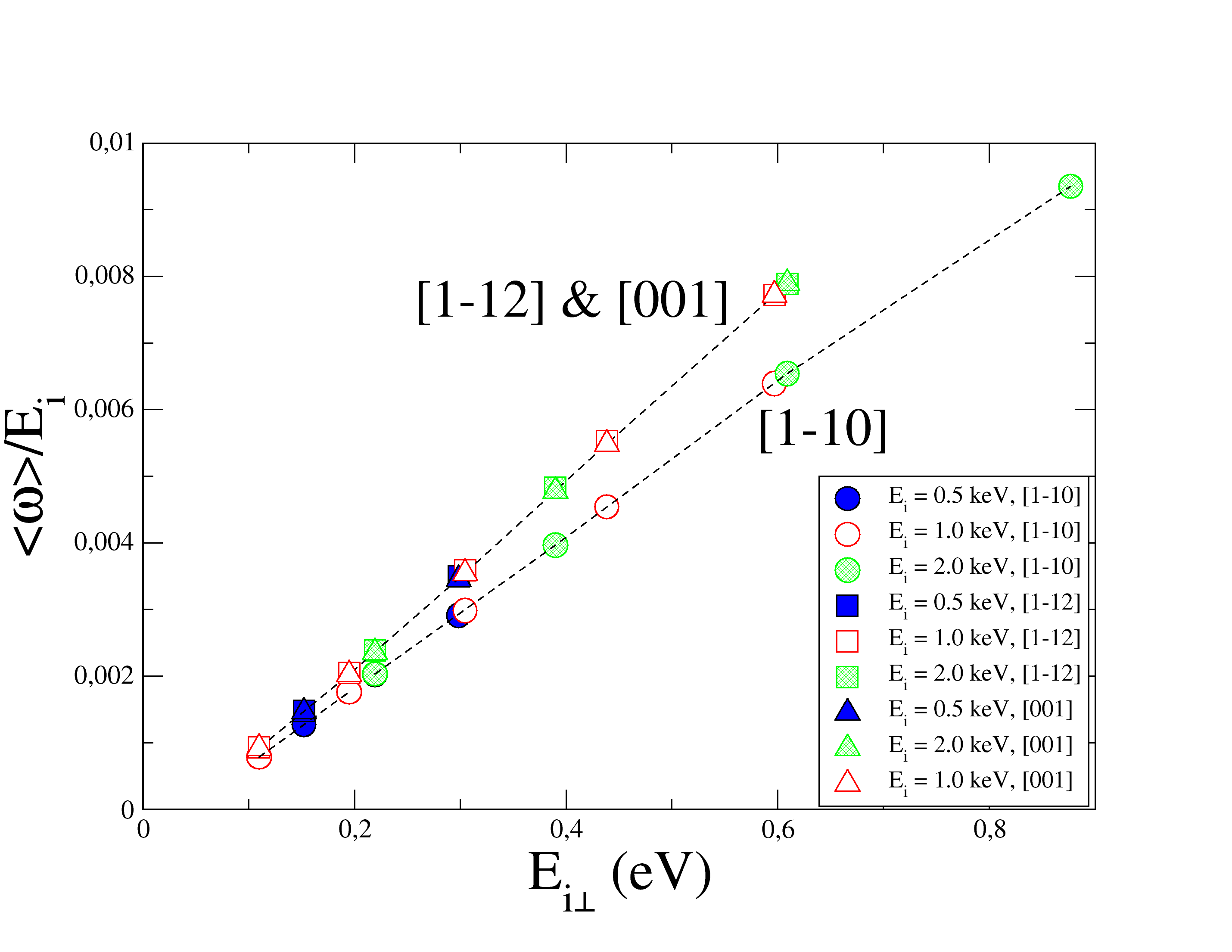}
\caption{(Color online) Normalized mean energy loss $\langle \protect\omega %
\rangle /E_{i}$, as a function of the perpendicular energy $E_{i\bot }$, for
different incidence energies and channels. Notation: circles, squares, and
triangles, results for incidence along the $[1\bar{1}0]$, $[1\bar{1}2]$, $%
[001] $ channels, respectively. Incidence energies according to the
following notation: full blue symbols, for $E_{i}=0.5$ keV; empty red
symbols, for $E_{i}=1.0$ keV; crossed green symbols, for $E_{i}=2.0$ keV.
Dashed lines are for guiding the eyes. }
\label{fig:7}
\end{figure}


\begin{thebibliography}{99}
\bibitem{note-conv} Theoretical results were convoluted with a Gaussian
function whose parameters were extracted from the experiment.
\end{thebibliography}
\end{document}